\documentstyle[11pt,moriondx,epsfig]{article}

\def\Journal#1#2#3#4{{#1} {\bf #2}, #3 (#4)}

\def\EPJC{{\em Eur.\ Phys.\ J.} C}
\def\CJP{\em Chin.\ J.\ Phys.}
\def\MPL{\em Mod.\ Phys.\ Lett.}

\def\NPB{{\em Nucl.\ Phys.} B}
\def\PLB{{\em Phys.\ Lett.}  B}
\def\PRL{\em Phys.\ Rev.\ Lett.}
\def\PRD{{\em Phys.\ Rev.} D}
\def\ZPC{{\em Z.\ Phys.} C}

\def\mevp{~MeV/$c$}
\def\mevm{~MeV/$c^2$}
\def\geve{~GeV}
\def\gevp{~GeV/$c$}
\def\gevm{~GeV/$c^2$}
\def\cherenkov{\v{C}erenkov}
\def\rmix{$r^{}_{\rm mix}$}
\def\cp{$CP$}

\def\pt{$p^{}_T$}
\def\ddbar{$D^0$-$\overline{D}^{\,0}$}
\def\simge{\mathrel{%
   \rlap{\raise 0.511ex \hbox{$>$}}{\lower 0.511ex \hbox{$\sim$}}}}
\def\simle{\mathrel{
   \rlap{\raise 0.511ex \hbox{$<$}}{\lower 0.511ex \hbox{$\sim$}}}}

\def\dkpi{$D^0\rightarrow K^-\pi^+$}
\def\dkk{$D^0\rightarrow K^- K^+$}
\def\dpipi{$D^0\rightarrow\pi^+\pi^-$}
\def\dppp{$D^+\rightarrow\pi^-\pi^+\pi^+$}
\def\dskpipi{$D^+_s\rightarrow K^-\pi^+\pi^+$}
\def\dkpipi{$D^+\rightarrow K^-\pi^+\pi^+$}
\def\kpipi{$K^-\pi^+\pi^+$}
\def\kll{$K^-\ell^+\ell^+$}
\def\dkll{$D^+\rightarrow K^-\ell^+\ell^+$}
\def\dkppp{$D^0\rightarrow K^-\pi^+\pi^-\pi^+$}
\def\dklnu{$D^0\rightarrow K^-\ell^+\nu^{}_{\ell}$}
\def\dkstarlnu{$D^+\rightarrow\overline{K}^{*0}\ell^+\nu^{}_{\ell}$}
\def\dphilnu{$D^+\rightarrow\phi\,\ell^+\nu^{}_{\ell}$}
\def\dskstark{$D^+_s\rightarrow\overline{K}^{*0}K^+$}
\def\dsphipi{$D^+_s\rightarrow\phi\,\pi^+$}
\def\dsphilnu{$D^+_s\rightarrow\phi\,\ell^+\nu^{}_{\ell}$}
\def\dstardpi{$D^{*+}\rightarrow D^0\pi^+$}
\def\dstardbarpi{$D^{*-}\rightarrow \overline{D}^{\,0}\pi^-$}

\begin{document}
\vspace*{0.5cm}
\title{
\begin{flushright}
{\large UCTP-112-99}
\end{flushright}
\vskip 0.15in
RECENT CHARM RESULTS FROM FERMILAB EXPERIMENT E791}

\author{ A.J. SCHWARTZ }

\address{Department of Physics, University of Cincinnati,\\
Cincinnati, Ohio 45221 (USA)}

\maketitle\abstracts{
Fermilab experiment E791 studied weak decays of $D^+$, $D^+_s$, and 
$D^0$ mesons produced in collisions of 500\gevp\ negative pions with
Pt and C targets. The experiment collected over 200\,000 fully 
reconstructed charm decays. Four recent results are discussed here:
{\it (a)}\ measurement of the form factor ratios $V/A^{}_1$, $A^{}_2/A^{}_1$, 
and $A^{}_3/A^{}_1$ in \dkstarlnu\ and \dsphilnu\ decays;
{\it (b)}\ measurement of the difference in decay widths $\Delta\Gamma$ 
between the two $D^0/\overline{D}^{\,0}$ mass eigenstates;
{\it (c)}\ search for rare and forbidden $D$ decays to dilepton 
final states; and
{\it (d)}\ search for a ``Pentaquark,'' a bound state of $\bar{c}suud$.
}

\section{Introduction}\label{sec:intro}
Fermilab E791 is a charm hadroproduction experiment studying
the weak decays of charmed mesons and baryons. The experiment
took data from September, 1991 to January, 1992, recording over 
$2\times 10^{10}$ interactions and reconstructing over 200\,000 
charm decays. This large sample has led to numerous published 
results.\,\cite{e791web} Four recent results are discussed here:
{\it (a)}\ measurement of the form factor ratios $V/A^{}_1$, $A^{}_2/A^{}_1$, 
and $A^{}_3/A^{}_1$ in \dkstarlnu\ and \dsphilnu\ decays;\,\cite{e791ff} 
{\it (b)}\ measurement of the difference in decay widths $\Delta\Gamma$  
between the two $D^0/\overline{D}^{\,0}$ mass eigenstates;\,\cite{e791deltgamm}
{\it (c)}\ search for rare and forbidden  $D$ decays;\,\cite{e791sanders} 
and 
{\it (d)}\ search for a ``Pentaquark,'' a bound state 
of $\bar{c}suud$.\,\cite{e791sharon,e791gilad}

The E791 collaboration comprises approximately 70 physicists 
from 17 institutions.\,\footnote{
C.B.P.F.\,(Brazil), Tel Aviv (Israel), CINVESTAV (Mexico), 
Puebla (Mexico), U.C.\,Santa Cruz, University of Cincinnati, 
Fermilab, Illinois Institute of Technology, Kansas State,
University of Massachusetts, University of Mississippi, 
Princeton, University of South Carolina, Stanford, Tufts, 
University of Wisconsin, and Yale.}
The experiment produced charmed mesons and baryons using a $\pi^-$ 
beam of momentum 
500\gevp\ incident on five thin target foils (one platinum, four carbon). 
The foils were separated along the beamline by approximately 1.5~cm such 
that most charm decays occurred in air rather than in solid material. 
Immediately downstream of the target was a silicon strip vertex detector 
consisting of 17 planes of silicon strips oriented along the $x,y,u$ and 
$v$ directions, where $u$ and $v$ point $\pm\,20^\circ$ from the vertical 
direction ($y$). Following the vertex detector was a spectrometer 
consisting of two large-aperture dipole magnets providing \pt\ kicks 
of 212\mevp\ and 320\mevp, and 37 planes of wire drift chambers and 
proportional chambers. Downstream of the second magnet were two threshold 
\cherenkov\ counters used to discriminate among pions, kaons, and protons. 
Following the \cherenkov\ counters was a Pb/liquid-scintillator
calorimeter used to measure the energy of electrons and photons, 
and an Fe/plastic-scintillator calorimeter used to measure the energy 
of hadrons. Downstream of the calorimeters was approximately 1.0~m of 
iron to range out any remaining hadrons, and following the iron were 
two stations of plastic scintillator --\,one $x$-measuring and one 
$y$-measuring\,-- to identify muons. The experiment used a loose 
transverse energy trigger ($E^{}_T\simge 3$\geve) that was almost 
fully efficient for charm decays.

After events were reconstructed, those with evidence of a decay vertex 
separated from the interaction vertex were retained for further analysis. 
The analyses presented here selected events using numerous kinematic and 
quality criteria; the most important of these are listed in 
Table~\ref{tab:cuts}. An especially effective criterion for enhancing 
signal over background was ``SDZ,'' the longitudinal distance between 
the production and decay vertices divided by the total measurement error 
in this quantity. 

\begin{table}
\caption{Kinematic and quality criteria used to select events. 
\label{tab:cuts}}
\vspace{0.4cm}
\begin{center}
\begin{tabular}{|l|c|}
\hline
{\bf Cut} & {\bf Typical value} \\
\hline
${\rm SDZ} \ =\  \left(z^{}_{\rm sec} - z^{}_{\rm prim}\right)/
       \sqrt{\sigma^2_{\rm sec}  + \sigma^2_{\rm prim}}$ & 
12 ($D^0,\,D^+_s$),\ \ 20 ($D^+$) \\
$p^{}_T$ (transverse to $D$ line-of-flight)  & $< 250$~MeV/$c$\\
$\Delta^{}_z ({\rm target}) = \left(z^{}_{\rm sec} - 
z^{}_{\rm targ.\ edge}\right)/\sigma^{}_{\rm sec}$  & $5.0 <$ \\
${\rm DIP}\ =\ $\,$D$ impact parameter w/r/t primary vertex
& $< 40~\mu$m \\
$\chi^2_{\rm track}$/(d.o.f.)  & $< 5.0$ \\
${\rm Lifetime}\ =\  \left(z^{}_{\rm sec} - z^{}_{\rm prim}\right)
		\cdot m^{}_D/p$ & 
$< 3.0$~ps ($D^0,\,D^+_s$),\ \ $< 5.0$~ps ($D^+$) \\
\hline		
\end{tabular}
\end{center}
\end{table}

\section{$D^+$ and $D^+_s$ Form Factors}\label{sec:formfactors}

Semileptonic decays such as \dkstarlnu\ and \dsphilnu\ proceed via 
spectator diagrams. As such, all hadronic effects are parametrized 
by four Lorentz-invariant form factors: $A_1(q^2)$, $A_2(q^2)$, $A_3(q^2)$, 
and $V(q^2)$. Unfortunately, the limited size of current data samples 
precludes measurement of the $q^2$ dependence, and we assume
this dependence to be given by a nearest pole dominance model: 
$F(q^2) = F(0)/(1-q^2/m^2_{\rm pole})$, where $m^{}_{\rm pole}=2.1$\gevm\ 
for the vector form factor $V$ and 2.5\gevm\ for the axial-vector form 
factors~$A$. Because $A_1(q^2)$ appears in every term in the differential 
decay rate, we factor out $A_1(0)$ and measure the ratios
$r^{}_V\equiv V(0)/A_1(0)$, $r^{}_2\equiv A_2(0)/A_1(0)$, and 
$r^{}_3\equiv A_3(0)/A_1(0)$. These ratios are insensitive to
the total decay rate and to the weak mixing matrix element $V^{}_{cs}$.

To select \dkstarlnu\ and \dsphilnu\ decays, we identify 3-track 
vertices in which one track is identified as a kaon and one track 
as a lepton. We cut on the ``transverse'' mass $m^{}_T$, where 
$m^2_T \equiv (E^{}_{K^*/\phi} + E^{}_\ell + p^{}_T)^2$\,$-$\,(
{\bf {\boldmath $p^{}_{K^*/\phi}$}}\,$+$\,{\bf {\boldmath $p^{}_\ell$}}\,$+$\, 
{\bf {\boldmath $p^{}_T$}})$^2$.
In this expression, {\bf {\boldmath $p^{}_T$}} is the momentum of the 
neutrino transverse to the direction of the $D$ as inferred by 
momentum balance. The $m^{}_T$ distribution of semileptonic decays forms 
a Jacobian peak with an endpoint at $m^{}_D$, and thus we require that
$m^{}_T$ lie in the range 1.6--2.0\gevm\ (1.7--2.1\gevm) for the
$D^+$ ($D^+_s$) sample. We also require that either 
$m^{}_{K\pi}\approx m^{}_{K^*}$ or $m^{}_{KK}\approx m^{}_\phi$. 
The resulting samples contain very little background, and we do 
a maximum likelihood fit for the form factors using a likelihood 
function based on three angles:
$\theta^{}_V$, the polar angle in the $\overline{K}^{*0}$
($\phi$) rest frame between the $\pi^+$ ($K^+$) and the $D^+$ ($D^+_s$);
$\theta^{}_\ell$, the polar angle in the $W^+$ rest frame between
the $\nu^{}_\ell$ and the $D^+$ ($D^+_s$); and
$\chi$, the azimuthal angle in the $D^+$ ($D^+_s$) rest frame
between the $\overline{K}^{*0}$ ($\phi$) and $W^+$ decay planes.
The results of the fit to the $\overline{K}^{*0}$ data for the
$e+\mu$ samples combined are: 
$r^{}_V = 1.87\,\pm\,0.08\,\pm\,0.07$ and
$r^{}_2 = 0.73\,\pm\,0.06\,\pm\,0.08$. 
We measure $r^{}_3 = 0.04\,\pm\,0.33\,\pm\,0.29$
from the $\mu$ sample alone.
The results for \dsphilnu\ are:
$r^{}_V = 2.27\,\pm\,0.35\,\pm\,0.22$ and 
$r^{}_2 = 1.57\,\pm\,0.25\,\pm\,0.19$. 
These results are compared with theoretical predictions in 
Table~\ref{tab:theory}; the errors in the measurements are 
smaller than the spread in theoretical predictions.
 
\begin{table}
\caption{Theoretical predictions for $r^{}_V$ and $r^{}_2$.
Some values are extrapolations from $q^2=q^2_{\rm max}$
to $q^2=0$. \label{tab:theory}}
\vspace{0.4cm}
\begin{center}
\begin{tabular}{|l|ll|}
\hline
{\bf Group} & {\boldmath {\bf $r^{}_V$}} &  {\boldmath {\bf $r^{}_2$}} \\
\hline
{\bf {\boldmath E791 \dkstarlnu}} &  
{\bf {\boldmath $1.87\,\pm\,0.11$}} & 
{\bf {\boldmath $0.73\,\pm\,0.10$ }} \\
\hline 
ISGW2\,\cite{isgw2}	& 2.0  		&	1.3		\\
WSG\,\cite{wsg}		&  1.4		&	1.3		\\
KS\,\cite{ks}		&  1.0		&	1.0		\\
AW/GS\,\cite{awgs}	&  2.0		&	0.8		\\
Stech\,\cite{stech}	&  1.55		&	1.06		\\
BKS\,\cite{bks}		&  $1.99\,\pm\,0.22\,\pm\,0.33$	& 
					$0.7\,\pm\,0.16\,\pm\,0.17$  \\
LMMS\,\cite{lmms}	&  $1.6\,\pm\,0.2$	& $0.4\,\pm\,0.4$ 	\\
ELC\,\cite{elc}		&  $1.3\,\pm\,0.2$	& $0.6\,\pm\,0.3$	\\
APE\,\cite{ape}		&  $1.6\,\pm\,0.3$	& $0.7\,\pm\,0.4$	\\
UKQCD\,\cite{ukqcd}	&  $1.4\,^{+0.5}_{-0.2}$	& $0.9\,\pm\,0.2$ \\
BBD\,\cite{bbd}		&  $2.2\,\pm\,0.2$	& $1.2\,\pm\,0.2$	\\
LANL\,\cite{lanl}	&  $1.78\,\pm\,0.07$	& $0.68\,\pm\,0.11$	\\
\hline
\hline
{\bf {\boldmath E791 \dsphilnu}} &  
{\bf {\boldmath $2.27\,\pm\,0.41$}} & 
{\bf {\boldmath $1.57\,\pm\,0.31$ }}  \\
\hline
ISGW2\,\cite{isgw2}	&  2.1  		&	1.3		\\
BKS\,\cite{bks}		&  $2.00\,\pm\,0.19\,^{+0.20}_{-0.25}$	& 
				$0.78\,\pm\,0.08\,^{+0.17}_{-0.13}$  \\
LMMS\,\cite{lmms}	&  $1.65\,\pm\,0.21$	& $0.33\,\pm\,0.33$ 	\\
\hline
\end{tabular}
\end{center}
\end{table}

\section{\ddbar\ Mixing and $\Delta\Gamma$}\label{sec:mixing}

E791 has published a limit on the \ddbar\ mixing rate using
semileptonic \dklnu\ decays\,\cite{e791arun} and hadronic \dkpi\ 
and \dkppp\ decays.\,\cite{e791hadromix} The flavor of the $D^0$ 
or $\overline{D}^{\,0}$ when produced is determined by combining the 
$D^0/\overline{D}^{\,0}$ with a low momentum pion to reconstruct
a \dstardpi\ or \dstardbarpi\ decay. The semileptonic decays 
yield a 90\% C.L.\ limit 
$r^{}_{\rm mix} < 0.50$\% [where $r^{}_{\rm mix}\equiv 
\Gamma(D^0\rightarrow\overline{D}^{\,0}\rightarrow\overline{f})/
\Gamma(D^0\rightarrow f)]$, while the hadronic decays  yield 
a 90\% C.L.\ limit $r^{}_{\rm mix} < 0.85$\%. 
The latter limit assumes no \cp\ violation in the mixing
and no \cp\ violation in a doubly-Cabibbo-suppressed 
(DCS) amplitude which also contributes to the rate. 
However, \cp\ violation is allowed in the {\it interference\/}
between the mixing and DCS amplitudes. Since the DCS amplitude 
is in fact substantially larger than that expected from mixing,
the presence of ``wrong-sign'' decays in the hadronic data 
--\,while a signature for mixing\,-- is more easily interpreted 
as evidence for DCS decays. If we assume no mixing, then the numbers 
of wrong-sign decays observed in our data, corrected for acceptance,
imply ratios of DCS decays to Cabibbo-favored decays of
$r^{K\pi}_{\rm DCS} = (0.68\,^{+0.34}_{-0.33}\,\pm\,0.07)$\% 
and $r^{K\pi\pi\pi}_{\rm DCS} = (0.25\,^{+0.36}_{-0.34}\,\pm\,0.03)$\%.
 
Since $r^{}_{\rm mix} = (1/2)\left[ (\Delta m/\Gamma)^2 + 
				(\Delta\Gamma/2\Gamma)^2\right]$,
where $\Delta m$ and $\Delta\Gamma$ are the differences between
the masses and decay widths of the $D^0/\overline{D}^{\,0}$ mass 
eigenstates, the upper limit for \rmix\ implies an upper limit 
for the difference in widths: $|\Delta\Gamma|< 0.48$~ps$^{-1}$. E791 
has  made a direct measurement of $\Delta\Gamma$ using \dkk\ and 
\dkpi\ decays. Since the former results in a \cp -even eigenstate, 
only the \cp -even component $D^0_1$ contributes and the lifetime 
distribution is proportional to $e^{-\Gamma^{}_1t}$. The $K^-\pi^+$ 
final state, however, is a \cp\ admixture and the lifetime 
distribution is proportional to 
$\exp\left[ -\left(\Gamma_1 + \Gamma_2\right) t/2\right]\,
			\cosh(\Delta\Gamma\,t/2)$.\,\cite{schwartz2}
Over the range of lifetimes for which the experiment has 
sensitivity, $\cosh(\Delta\Gamma\,t/2)\approx 1$ and thus:
$\Gamma^{}_{KK} - \Gamma^{}_{K\pi} 
= \Gamma^{}_1 - (\Gamma^{}_1 + \Gamma^{}_2)/2  =
(\Gamma^{}_1 - \Gamma^{}_2)/2 = \Delta\Gamma/2$.

\begin{figure}
\hbox{
\mbox{\epsfig{figure=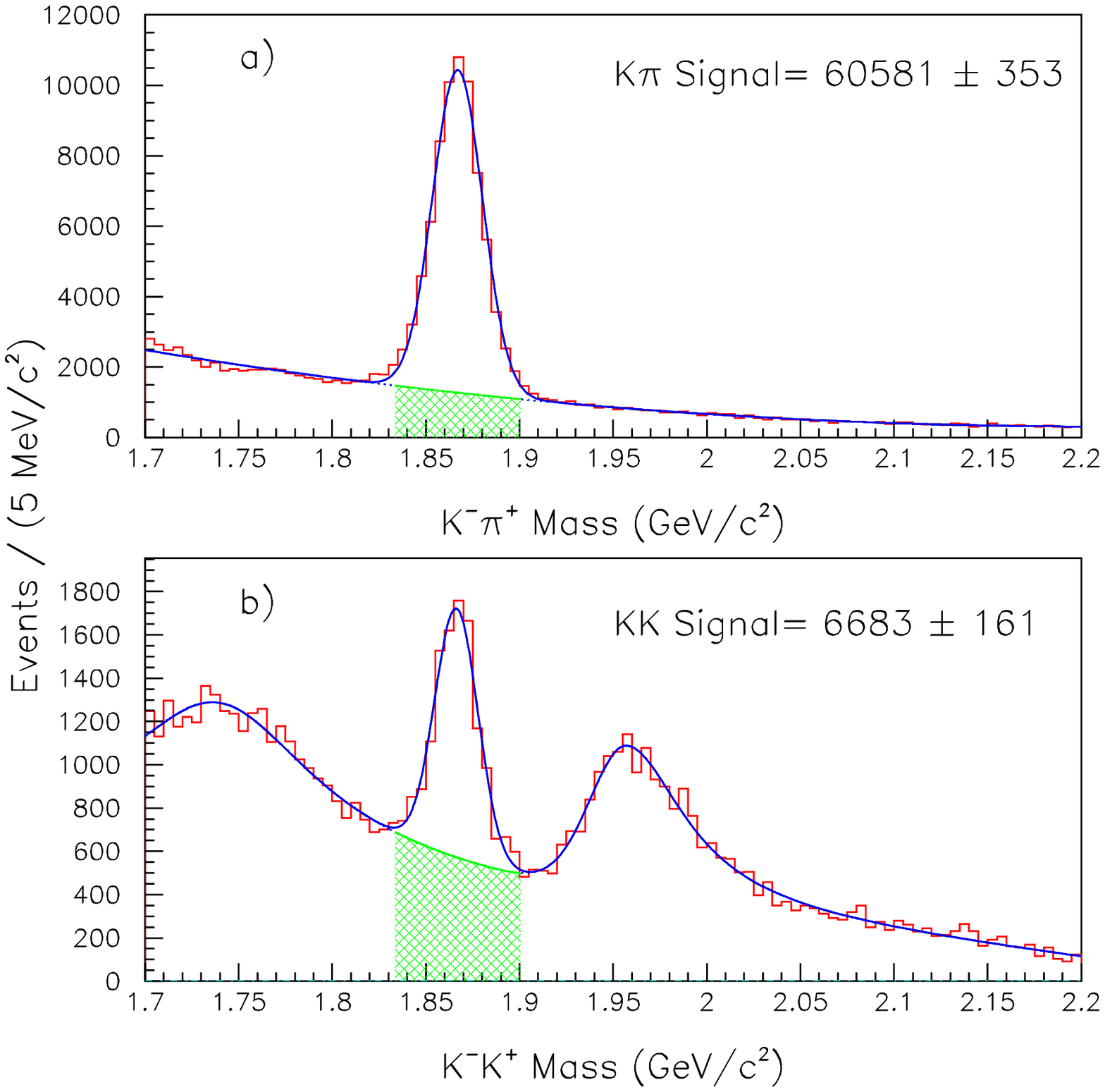,height=2.8in}}
\hspace*{0.15in}
\mbox{\epsfig{figure=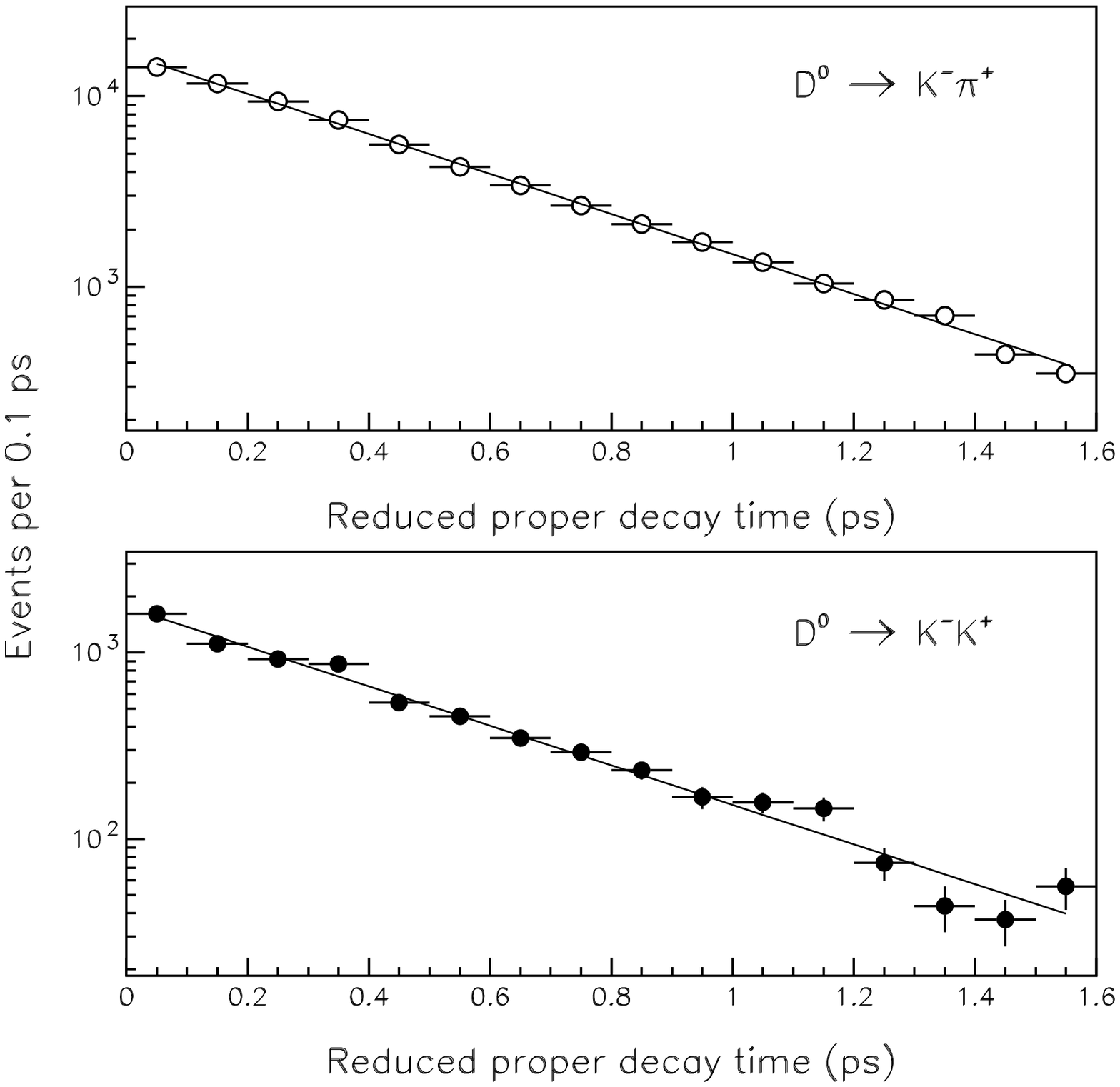,height=2.8in}}
}
\caption{\dkpi\ and \dkk\ mass distributions (left), and reduced 
proper lifetime distributions (right). The right-most peak in the 
lower left plot is due to misidentified \dkpi\ decays.
\label{fig:kkpp}}
\end{figure}

Our samples of \dkk\ and \dkpi\ are shown in Fig.~\ref{fig:kkpp}a.
We bin these events by {\it reduced\/} proper lifetime, which is the 
distance traveled by the $D^0$ candidate beyond that required to survive 
our selection criteria, multiplied by mass and divided by momentum.
For each bin of reduced lifetime we fit the mass distribution for the 
number of signal events. Plotting this number (corrected for acceptance)
as a function of reduced lifetime gives the distributions shown in 
Fig.~\ref{fig:kkpp}b. Fitting these distributions to exponential 
functions yields $\Gamma^{}_{KK} = 2.441\,\pm\,0.068$~ps$^{-1}$, 
$\Gamma^{}_{K\pi} = 2.420\,\pm\,0.019$~ps$^{-1}$, and thus
$\Delta\Gamma = 0.04\,\pm\,0.14\,\pm\,0.05$~ps$^{-1}$. This implies 
$-0.20<\Delta\Gamma < 0.28$~ps$^{-1}$ at 90\% C.L., which is more 
stringent than the constraint resulting from  $r^{}_{\rm mix}$.

\section{Rare and Forbidden $D$ Decays}\label{sec:fcnc}

E791 has searched for rare and forbidden dilepton decays of the
$D^+$, $D^+_s$, and $D^0$. The decay modes can be classified as follows:
\begin{enumerate}
\item flavor-changing neutral current decays $D^0\rightarrow\ell^+\ell^-$
and $D^+_{(d,s)}\rightarrow h^+\ell^+\ell^-$, in which $h$ is a pion or 
kaon;
\item lepton-flavor violating decays $D^0\rightarrow\mu^\pm e^\mp$, 
$D^+_{(d,s)}\rightarrow h^+\mu^\pm e^\mp$, and 
$D^+_{(d,s)}\rightarrow h^-\mu^+ e^+$, in which the leptons belong to 
different generations; and
\item lepton-number violating decays 
$D^+_{(d,s)}\rightarrow h^-\ell^+\ell^+$, in which the leptons 
belong to the same generation but have the same sign charge.
\end{enumerate}
Decay modes belonging to (1) occur within the Standard Model 
via higher-order diagrams, but the branching fractions are 
estimated\,\cite{schwartz} to be only $10^{-8}$ to $10^{-6}$. 
This is below the sensitivity of current experiments. Decay modes 
belonging to (2) and (3) do not conserve lepton number and thus 
are forbidden within the Standard Model. However, a number of 
theoretical extensions to the Standard Model predict lepton 
number violation,\,\cite{lnvpredict} and the observation of 
a signal in these modes would indicate new physics. We have 
searched for 24 different rare and forbidden decay modes and have 
found no evidence for them. We therefore present upper limits on 
their branching fractions. Eight of these modes have no previously 
reported limits, and fourteen are reported with substantial 
improvements over previously published results.

For this study we used a ``blind'' analysis technique. Before our 
selection criteria were finalized, all events having masses within a 
window $\Delta M_S$ around the mass of the $D^+$, $D^+_s$, or $D^0$ were 
masked so that the presence or absence of potential signal candidates 
would not bias our choice of selection criteria. All criteria were then 
chosen by studying signal events generated by Monte Carlo simulation and
background events obtained from the data. The background events were 
chosen from 
mass windows $\Delta M_B$ above and below the signal window $\Delta M_S$. 
The criteria were chosen to maximize the ratio $N^{}_S/\sqrt{N^{}_B}$, 
where $N^{}_S$ and $N^{}_B$ are the numbers of signal and background 
events, respectively. Only after this procedure were events within the 
signal window unmasked. The signal windows $\Delta M_S$ used for decay 
modes containing electrons are asymmetric around $m^{}_D$ to allow for 
the bremsstrahlung low-energy tail. 

We normalize the sensitivity of our search to topologically similar 
Cabibbo-favored decays. For the $D^+$ modes we use \dkpipi; for the 
$D^+_s$ modes we use \dsphipi; and for the $D^0$ we use \dkpi. The 
upper limit on the branching fraction for decay mode $X$ is:
\begin{equation}
B^{}_X = \frac{N^{}_X}{N^{}_{\rm norm}}\,
\frac{\varepsilon^{}_{\rm norm}}{\varepsilon^{}_X} \cdot B^{}_{\rm norm}\ ,
\end{equation} 
where $N^{}_X$ is the upper limit on the mean number of signal events,
$N^{}_{\rm norm}$ is the number of normalization events,
and $\varepsilon^{}_X$ and $\varepsilon^{}_{\rm norm}$ are 
overall detection efficiencies. 
The geometric acceptances and reconstruction efficiencies 
are found from Monte Carlo simulation, and the particle 
identification efficiencies are measured from data. 

\begin{figure}[tb]
\begin{center}
\mbox{\epsfig{figure=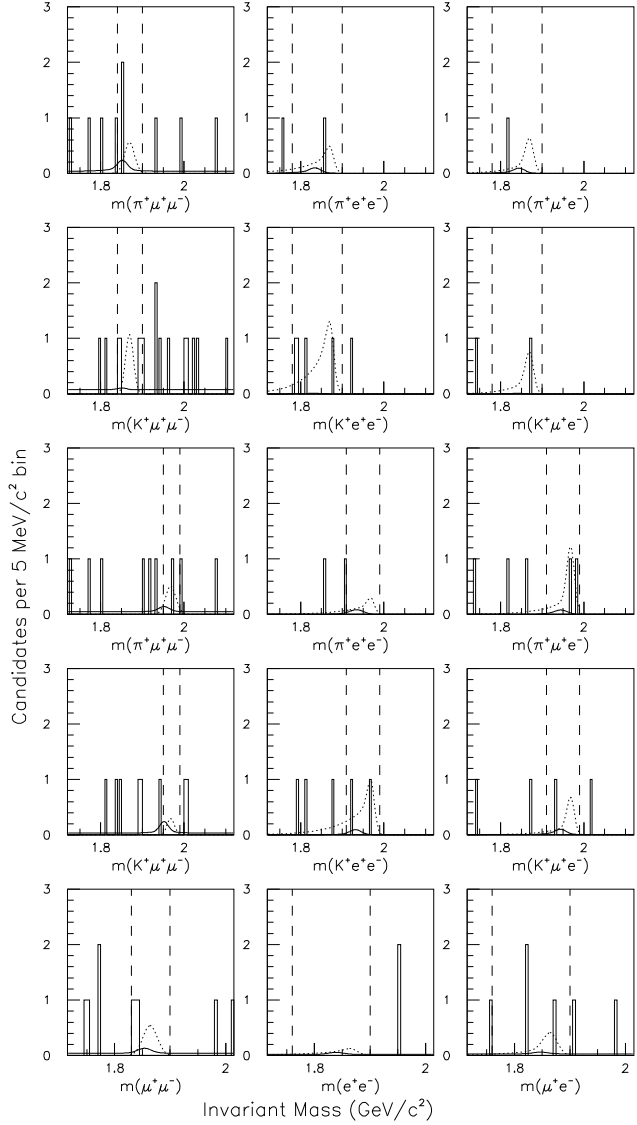,height=6.0in}}
\end{center}
\caption{Final event samples for $D^+$ (rows 1--2), 
$D^+_s$ (rows 3--4), and $D^0$ (row 5) decays. The dotted
curves show signal shape for a number of events equal to 
the 90\% C.L.\ limit. The solid curves show estimated 
background. \label{fig:sanders_mass}}
\end{figure}

The background consists of random combinations of tracks and vertices, 
and reflections from more copious hadronic $D$ decays. The former is 
essentially flat in the reconstructed invariant mass, and we estimate 
this background by scaling the level from mass regions above and below 
the signal region $\Delta M_S$. The hadronic decay background in which 
a $K$ is misidentified as a lepton is explicitly removed via a 
$K\pi\pi$ or $KK\pi$ invariant mass cut. The hadronic background 
in which a $\pi$ is misidentified as a lepton cannot be removed 
in this manner, as the reflected mass and true mass are too close 
and such a cut would remove a substantial fraction of signal events. 
We thus estimate this background by multiplying the number of 
\dppp, \dskpipi, or \dpipi\ decays falling within the signal region 
$\Delta M_S$ by the rate for double particle misidentification
$\pi\pi\rightarrow\mu\mu,\ \mu e$, or $ee$. The misidentification
rates were measured from data using \dkpipi\ decays misidentified 
as \kll. Because the latter samples have substantial feedthrough
background from the former (which is Cabibbo-favored), we do not 
attempt to establish a limit for \dkll\ decays. Rather, we use 
the observed signals to measure the lepton misidentification rates 
under the assumption that all \kll\ decays observed arise from 
misidentified \kpipi. Most of our final event samples are shown 
in Fig.~\ref{fig:sanders_mass}, and all results are tabulated in 
Table~\ref{tab:fcncresults}.

\section{Search for the Pentaquark $P^0_{\bar{c}suud}$}\label{sec:penta}

E791 has searched for a ``Pentaquark'' $P^0$, which is a bound state
of five quarks having flavor quantum numbers $\overline{c}suud$. 
This state was originally proposed by Lipkin\,\cite{lipkin} and
Gignoux {\it et al.}\,\cite{gignoux} over ten years ago, but no
experimental searches have been undertaken. The $P^0$ is predicted
to have a mass below the threshold for strong decay  
($m^{}_{D_s} + m^{}_{p} = 2.907$\gevm) by 10--150\mevm. The lifetime 
is expected to be similar to that of the shortest-lived charm meson, 
0.4--0.5~ps. We have searched for $P^0$ decays into $\phi\,p\,\pi^-$ 
and $K^{*0}K^- p$\/ final states.

The analysis proceeds by first selecting four-track vertices in which 
one track is identified as a proton and two opposite-sign tracks 
are identified as kaons. We require that either 
$m^{}_{KK}\approx m^{}_\phi$ or $m^{}_{K\pi}\approx m^{}_{K^*}$ 
and remove events in which either $m^{}_{p\pi}\approx m^{}_\Lambda$ 
or the $\phi$ or $p$ momentum projects back to the production vertex. 
We normalize the sensitivity of the search to \dsphipi\ and \dskstark\ 
decays; these are topologically similar to $P^0\rightarrow\phi\,p\,\pi^-$ 
and $P^0\rightarrow K^{*0}K^- p$ (except for the proton) and several 
systematic errors cancel.
After all selection criteria are applied, we observe no excess 
of events above background in either decay channel. We thus obtain 
upper limits for the product of production cross section and 
branching fraction, relative to that for the $D^+_s$. The 
expression used is (here for $\phi\,p\,\pi^-$):
\begin{equation}
\frac{\sigma^{}_P\cdot B^{}_{P\rightarrow\phi p \pi}}
     {\sigma^{}_{D^{}_s}\cdot B^{}_{D^{}_s\rightarrow\phi\pi}}  
\ =\ \frac{N^{}_{P\rightarrow\phi p \pi}}{N^{}_{D^{}_s\rightarrow\phi\pi}}
\ \frac{\varepsilon^{}_{D^{}_s\rightarrow\phi\pi}}
{\varepsilon^{}_{P\rightarrow\phi p\pi}}\ ,
\label{eqn:pentalimit}
\end{equation}
where $N^{}_{P\rightarrow\phi\,p \pi}$ is the upper 
limit on the mean number of $P^0\rightarrow\phi\,p \pi$ decays, and
$N^{}_{D^{}_s\rightarrow\phi\,\pi}$ is the number of events observed
in the normalization channel. All numbers and the resulting limits are 
listed in Table~\ref{tab:sharon}. When calculating acceptance, we assume 
the $P^0$ lifetime to be 0.4~ps. The limits are given for two possible 
values of $m^{}_{P^0}$; the difference in the limits is due mainly to 
the difference in the numbers of events observed in the mass spectrum 
around these mass values. Our upper limits are 2--4\% of that for the 
corresponding $D^+_s$ decay, which is similar to the theoretical 
estimate ($\sim$\,1\%).

\begin{table}
\caption{90\% C.L.\ upper limits for dilepton decays of 
$D^+/D^+_s/D^0$. The right-most column lists previous results. 
\label{tab:fcncresults}}
\vspace{0.4cm}
\begin{center}
\begin{tabular}{|l|cccll|}
\hline
& & & & & \\
{\bf Mode} & {\bf Background} & {\bf Observed} & {\boldmath {\bf UL ($N$)}}
  & {\boldmath {\bf UL($B$)}} & {\boldmath {\bf PDG\,98}}\,\cite{pdg}  \\
  &   &  &  &	{\boldmath {\bf $(\times 10^5)$}} &   
		{\boldmath {\bf $(\times 10^5)$}}  \\
\hline
$D^+\rightarrow\pi^+\mu^+\mu^-$  &  2.67  &  2  &  3.35  & $<1.5$ & $<1.8$ \\
$D^+\rightarrow\pi^+ e^+ e^-$    &  0.90  &  1  &  3.53  & $<5.2$ & $<6.6$ \\
$D^+\rightarrow\pi^+\mu^{\pm}e^{\mp}$  
				 &  0.78  &  1  &  3.64  & $<3.4$ & $<12$ \\
& & & & & \\
$D^+\rightarrow\pi^-\mu^+\mu^+$  &  1.53  &  1  &  2.92  & $<1.7$ & $<8.7$ \\
$D^+\rightarrow\pi^- e^+ e^+$    &  0.45  &  2  &  5.60  & $<9.6$ & $<11$ \\
$D^+\rightarrow\pi^-\mu^+e^+$    &  0.39  &  1  &  4.05  & $<5.0$ & $<11$ \\
& & & & & \\
$D^+\rightarrow K^+\mu^+\mu^-$  &  2.40  &  3  &  5.07  & $<4.4$ & $<9.7$ \\
$D^+\rightarrow K^+ e^+ e^-$    &  0.09  &  4  &  8.72  & $<20$ & $<20$ \\
$D^+\rightarrow K^+\mu^{\pm}e^{\mp}$  
				 &  0.08 &  1  &  4.34  & $<6.8$ & $<13$ \\
& & & & & \\
$D^+_s\rightarrow K^+\mu^+\mu^-$  &  2.00  &  0  &  1.32  & $<14$ & $<59$ \\
$D^+_s\rightarrow K^+ e^+ e^-$    &  0.85  &  2  &  5.77  & $<160$ & $\ \ -$ \\
$D^+_s\rightarrow K^+\mu^{\pm}e^{\mp}$  
				  &  1.10  &  1  &  3.57  & $<63$ & $\ \ -$ \\
& & & & & \\
$D^+_s\rightarrow K^-\mu^+\mu^+$  &  1.04  &  0  &  1.68  & $<18$ & $<59$ \\
$D^+_s\rightarrow K^- e^+ e^+$    &  0.39  &  0  &  2.22  & $<63$ & $\ \ -$ \\
$D^+_s\rightarrow K^-\mu^+e^+$    &  1.15  &  1  &  3.53  & $<68$ & $\ \ -$ \\
& & & & & \\
$D^+_s\rightarrow\pi^+\mu^+\mu^-$  &  1.65  &  1  &  3.02  & $<14$ & $<43$ \\
$D^+_s\rightarrow\pi^+ e^+ e^-$    &  0.83  &  0  &  1.85  & $<27$ & $\ \ -$ \\
$D^+_s\rightarrow\pi^+\mu^{\pm}e^{\mp}$  
				  &  0.72  &  2  &  6.01  & $<61$ & $\ \ -$ \\
& & & & & \\
$D^+_s\rightarrow\pi^-\mu^+\mu^+$  &  1.16  &  0  &  1.60  & $<8.2$ & $<43$ \\
$D^+_s\rightarrow\pi^- e^+ e^+$    &  0.42  &  1  &  4.44  & $<69$ & $\ \ -$ \\
$D^+_s\rightarrow\pi^-\mu^+e^+$    &  0.36  &  3  &  8.21  & $<73$ & $\ \ -$ \\
& & & & & \\
$D^0\rightarrow \mu^+\mu^-$	  &  2.46  &  2  &  3.51  & $<0.52$ & $<0.41$
 \\
$D^0\rightarrow e^+e-$		  &  2.04  &  0  &  1.26  & $<0.62$ & $<1.3$ \\
$D^0\rightarrow \mu^{\pm}e^{\mp}$ &  2.88  &  2  &  3.09  & $<0.81$ & $<1.9$ \\
\hline
\end{tabular}
\end{center}
\end{table}

\begin{table}
\caption{90\% C.L.\ upper limits on the product of
cross section and branching fraction ($\sigma\cdot B$) for
$P^0\rightarrow\phi p\pi^-$ decays (left table) and
$P^0\rightarrow K^{*0}K^- p$ decays (right table),
relative to that for \dsphipi\ and \dskstark. \label{tab:sharon}}
\vspace{0.4cm}
\hbox{
\begin{tabular}{|l|cc|}
\hline
  & \multicolumn{2}{c|}{\bf {\boldmath $P^0$ mass (GeV/$c^2$)}} \\
  &   {\bf 2.79} & {\bf 2.87} \\
\hline
$UL(N^{}_{P^0\rightarrow\phi p\pi})$: & 3.1 & 7.5 \\
$\varepsilon^{}_P/\varepsilon^{}_{D^{}_s}$: & 0.47 & 0.64 \\
$N^{}_{D^{}_s}$:   &  \multicolumn{2}{c|}{$293\,\pm\,18$} \\
\hline
Upper limit: & 0.022 & 0.040  \\
\hline
\end{tabular}
\hspace*{0.60in}
\begin{tabular}{|l|cc|}
\hline
  & \multicolumn{2}{c|}{\bf {\boldmath $P^0$ mass (GeV/$c^2$)}} \\
  &   {\bf 2.77} & {\bf 2.85} \\
\hline
$UL(N^{}_{P^0\rightarrow K^{*}Kp})$: & 6.1 & 3.5 \\
$\varepsilon^{}_P/\varepsilon^{}_{D^{}_s}$: & 0.23 & 0.31 \\
$N^{}_{D^{}_s}$:   &  \multicolumn{2}{c|}{$725\,\pm\,88$} \\
\hline
Upper limit: & 0.036 & 0.016  \\
\hline
\end{tabular}
}
\end{table}

\section{Summary}\label{sec:summary}

We have presented four recent results from Fermilab experiment E791:
a measurement of the form factors governing \dkstarlnu\ and 
\dphilnu\ decays;
a measurement of the difference in decay widths $\Delta\Gamma$  
between the two mass eigenstates of $D^0/\overline{D}^{\,0}$; 
new limits on two dozen rare and forbidden dilepton decays
of $D^0$, $D^+$, and $D^+_s$; and
a limit on $\sigma\cdot B$ for a ``Pentaquark'' $P^0$ 
relative to that for the $D^+_s$.
Almost all measurements and limits are superior to previously 
published results. In the case of the $P^0$ and eight of the 
rare and forbidden dilepton decays, our limits are the first 
such limits reported.

\end{document}